\documentclass [amssymb,amsmath,onecolumn, showpacs]{revtex4-1} 
\usepackage{graphicx,epsfig,amsfonts,amssymb}
\usepackage{bm}
\usepackage{times}
\usepackage{lipsum}
\usepackage{verbatim}

\newcommand{\be}{\begin{equation}}
\newcommand{\ee}{\end{equation}}
\newcommand{\bes}{\begin{subequations}}
\newcommand{\ees}{\end{subequations}}
\newcommand{\bea}{\begin{eqnarray}}
\newcommand{\eea}{\end{eqnarray}}
\newcommand{\ba}{\begin{array}}
\newcommand{\ea}{\end{array}}
\newcommand{\beqn}{\begin{eqnarray*}}
\newcommand{\eeqn}{\end{eqnarray*}}

\newcommand{\la}{\langle}
\newcommand{\ra}{\rangle}

\newcommand{\mP}{\mathcal{P}}

\newcommand{\rar}{\rightarrow}

\newcommand{\nn}{\nonumber}
\begin{document}

\title{Exact transition probabilities for a linear sweep through a Kramers-Kronig resonance}

\author {Chen Sun$^{a,b}$ and N.~A. {Sinitsyn}$^{a}$ }
\address{$^a$ Theoretical Division, Los Alamos National Laboratory, Los Alamos, NM 87545,  USA}
\address{$^b$ Department of Physics, Texas A\&M University, TX 77840,  USA}

\begin{abstract}
We consider a localized electronic spin controlled by a circularly polarized optical beam and an external magnetic field. When the frequency of the beam is tuned near an optical resonance with a continuum of higher energy states, effective magnetic fields are induced on the two-level system via the Inverse Faraday Effect. We explore the process in which the frequency of the beam is made linearly time-dependent so that it sweeps through the optical resonance, starting and ending at the values far away from it. In addition to changes of spin states, Kramers-Kronig relations guarantee that a localized electron can also escape into a continuum of states. We argue that probabilities of transitions between different possible electronic states after such a  sweep of the optical frequency can be found exactly regardless the shape of the resonance. We also discuss extension of our results to multistate systems.
\end{abstract}
\date{\today}

\maketitle

\section{Introduction}

Light interaction with a magnetic system is sensitive to optical polarization. For example, absorption of a circularly polarized light in a semiconductor depends on the magnetic moment of conduction electrons \cite{tmd}. In addition to such dissipative effects, a polarized optical beam can experience more subtle magneto-optical effects, such as rotation of the linear polarization axis of a beam, called the Faraday Rotation. It is frequently used in
optical spectroscopy \cite{crooker}. A related effect is the shift of the phase of the circularly polarized light passing through a medium with spin polarized electrons \cite{phase-shift}.
Microscopically, such effects are described by virtual dissipationless excitations of the magnetic degrees of freedom.
Despite seemingly different microscopic origin of dissipative and nondissipative magneto-optical effects, causality of the magnetic medium response to the interactions with light imposes constraints between these effect. Such constraints are known as the Kramer-Kronig relations \cite{causality}.

As magnetic media influence the polarized light, there is a feedback from this interaction on the dynamics of the electron spin polarization. Moreover, both dissipative  changes of light intensity and dissipationless phase shifts of the beam induce counterpart effects on electronic spins, such as dissipative optically induced electronic spin polarization in semiconductors \cite{tmd} and dissipationless appearance of an effective, optically induced, magnetic fields acting on electronic spins. The latter effect is generally known as the Inverse Faraday Effect \cite{inv-far}.
It  has attracted considerable interest recently due to applications to purely optical ultrafast control of magnetization \cite{inv-far-exp}. The field of research called the circuit QED  \cite{girvin-QED} aims in engineering such light  interactions with discrete quantum states in order to realize desired entanglement and dynamics of solid state qubits.

Coupling of quantum spin systems to polarized light can be strongly enhanced when the frequency of light is tuned near an optical resonance with excited states of electrons. However, unlike the application of time-dependent fields, Kramers-Kronig relations guarantee that in addition to effective time-dependent fields, optical pulses inevitably induce dissipative transitions \cite{awschalom} that break coherence or even remove electrons, e.g., from a confining potential of quantum dots. Hence, in order to optimize the optical qubit control, it is important to develop the theory of  qubit interactions with time-depenent optical pulses at near resonant conditions.

In this article, we derive an exact nonperturbative result, which is akin to the famous Landau-Zener formula \cite{book}. The latter is the solution of the model that describes interaction of a two-level system  with a linearly time-dependent magnetic field. This model is used frequently in applications of nonstationary quantum mechanics, e.g., for characterization of parameters of nanoscale systems \cite{LZ-interferometry,app-spin,coher} and developing quantum control strategies \cite{qcontrol,recent-MLZ}.

In our case, instead of a linearly time-dependent  magnetic field, we consider a situation when a spin  system interacts with a polarized optical beam such that the frequency of the beam changes linearly with time and sweeps through an optical resonance. As in the case of the Landau-Zener model, when the frequency of the beam is far away from the resonance, interactions of the beam with the two-level system are negligible. Only during a finite time interval, when the frequency of the beam is close to the resonant frequency, can the beam induce additional dynamics of a two-level system. Hence, one can define a scattering problem: given the state of the spin system and $t\rightarrow - \infty$, what is the state of the spin at $t \rightarrow +\infty$? Using the approach pioneered by Landau \cite{landau} and some of the recent developments \cite{sinitsyn-14pra}, we demonstrate that one can derive simple exact analytical expressions for transition probabilities between spin states in this scattering process.

\section{The Model}

We consider a situation shown in Fig.~\ref{setup-fig}. Circularly polarized light interacts with spin up  (but not down) state of a two-level system.
The frequency of the light $\omega$ is tuned to be near the resonant frequency $\omega_0$ of a transition to a continuum of high energy states. If such a transition happens irreversibly, electron is lost in the continuum. In addition to such irreversible transitions, electron can produce virtual coherent transitions that merely renormalize the energy of the spin up state.

One can describe the interaction of a spin-1/2 with the light plus a constant external magnetic field ${\bf h}$ by the Schr\"odinger equation
\be
i\frac{d}{dt}\psi = \hat{H} (t) \psi
\label{sch1}
\ee
with an effective Hamiltonian:
\begin{figure}
\scalebox{0.15}[0.15]{\includegraphics{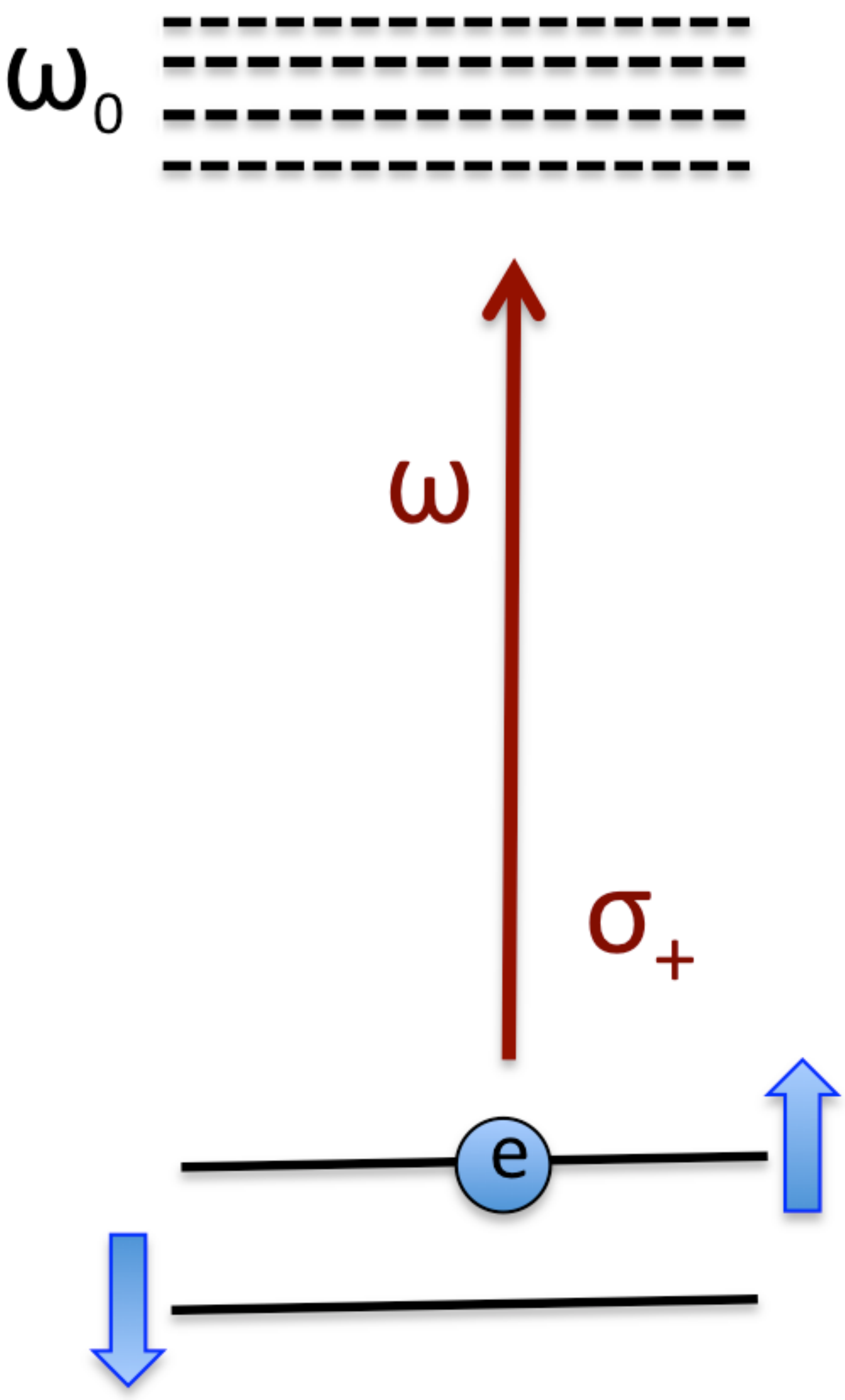}}
\hspace{-2mm}\vspace{-4mm}
\caption{ (Color online) An electron with spin 1/2 is initially in  a superposition of two spin states with energies split by a magnetic field. A circularly polarized beam with frequency $\omega$ interacts with this electron. The optical frequency $\omega$ is tuned near the resonant frequency $\omega_0$ that corresponds to the excitation of an electron into a continuum of much higher energy states. We assume that if an irreversible transition in the continuum happens, the electron has negligible probability to return to the initial doublet during the time of experiment. Frequency of the beam is assumed to be linearly time-dependent $\omega-\omega_0=\beta t$, sweeping from large negative to large positive $t$ with a constant rate $\beta>0$.  Circularly polarized field has nonzero matrix elements between the states of the continuum and the spin ``up" state.
Spin ``down" state does not interact with this field.
}
\label{setup-fig}
\end{figure}
\begin{equation}
\hat{H} = \left( \begin{array}{cc}
\chi(\omega) & 0 \\
0& 0
\end{array}\right) + {\bf h \cdot \hat{\sigma}},
\label{ham}
\end{equation}
 where $\chi(\omega)$ is the response function that describes renormalization of the energy of the spin up sate. This function is complex valued due to the possibility of irreversible excitations. Let us explicitly introduce the real part of this function $\chi_1(\omega)$ and its imaginary part $\chi_2(\omega)$:
  \begin{equation}
 \chi(\omega)=\chi_1(\omega) + i \chi_2(\omega).
 \label{chi1}
 \end{equation}
  Karamers-Kronig constraints relate them as
  \begin{eqnarray}
  \label{kk-1}
 \chi_1(\omega) =\frac{1}{\pi} \mP \int_{-\infty}^{\infty} \frac{\chi_2(\omega')} {\omega' - \omega} d\omega' ,\\
 \chi_2(\omega) =-\frac{1}{\pi} \mP \int_{-\infty}^{\infty} \frac{\chi_1(\omega')} {\omega' - \omega} d\omega',
 \label{kk-2}
 \end{eqnarray}
where the symbol $\mP$ near the integral means that only the principle part of this integral is considered.

In what follows, we assume an almost arbitrary form of the function $\chi(\omega)$ except that its dissipative part is localized in the frequency domain, so that the integral
\be
k=-\frac{1}{\pi} \int d\omega' \chi_2(\omega')
\label{qq}
\ee
 is finite. The constant $k$  characterizes the strength of the resonance.

 In appendix A we work out a specific  case of a broad resonance with a weak beam intensity, for which expressions for $\chi_1(\omega)$ and $\chi_2(\omega)$ can be obtained explicitly, and Kramers-Kronig relations can be verified.
 It is  important to note here that  the Kramers-Kronig relations can be justified  physically in the context of Eq.~(\ref{sch1}) only if dissipative electronic transitions are truly irreversible, which means that being excited to the continuum, the electron must leave the system. Such an experimental situation can be found  in the literature, e.g., on spin decoherence in InGaAs quantum dots \cite{alex-nat}. On the other hand, if the excited electron can return to the quantum dot during the time of experiment, one should use a more complicated density matrix approach, to which our discussion does not apply.
In what follows, we will focus on mathematical properties of the dynamics described by equations (\ref{sch1})-(\ref{qq}).

 The relation between causality, analyticity and Karamers-Kronig relations, which is discussed thoroughly, e.g., in \cite{causality}, predicts that any function $\chi(\omega)$ that satisfies Eqs.~(\ref{kk-1})-(\ref{kk-2}) must be analytic function of $\omega$ in the upper-half of the complex plane (Re($\omega$),Im($\omega$)). 
Setting $\omega \rar \infty$, we can disregard $\omega'$ in denominator  of (\ref{kk-1}) and find that
 \be
 \chi_1(\omega) \sim \frac{k}{\omega-\omega_0}, \quad |\omega-\omega_0|\gg \gamma,
 \label{asympt1}
 \ee
where $\gamma$ is a characteristic width of the resonance in the frequency space, and $k$ is a constant given by Eq.~(\ref{qq}). Thus we find that the real part of $\chi(\omega)$  decays as $\sim 1/\omega$ at $\omega \rightarrow \infty$.

\begin{figure}
\scalebox{0.5}[0.5]{\includegraphics{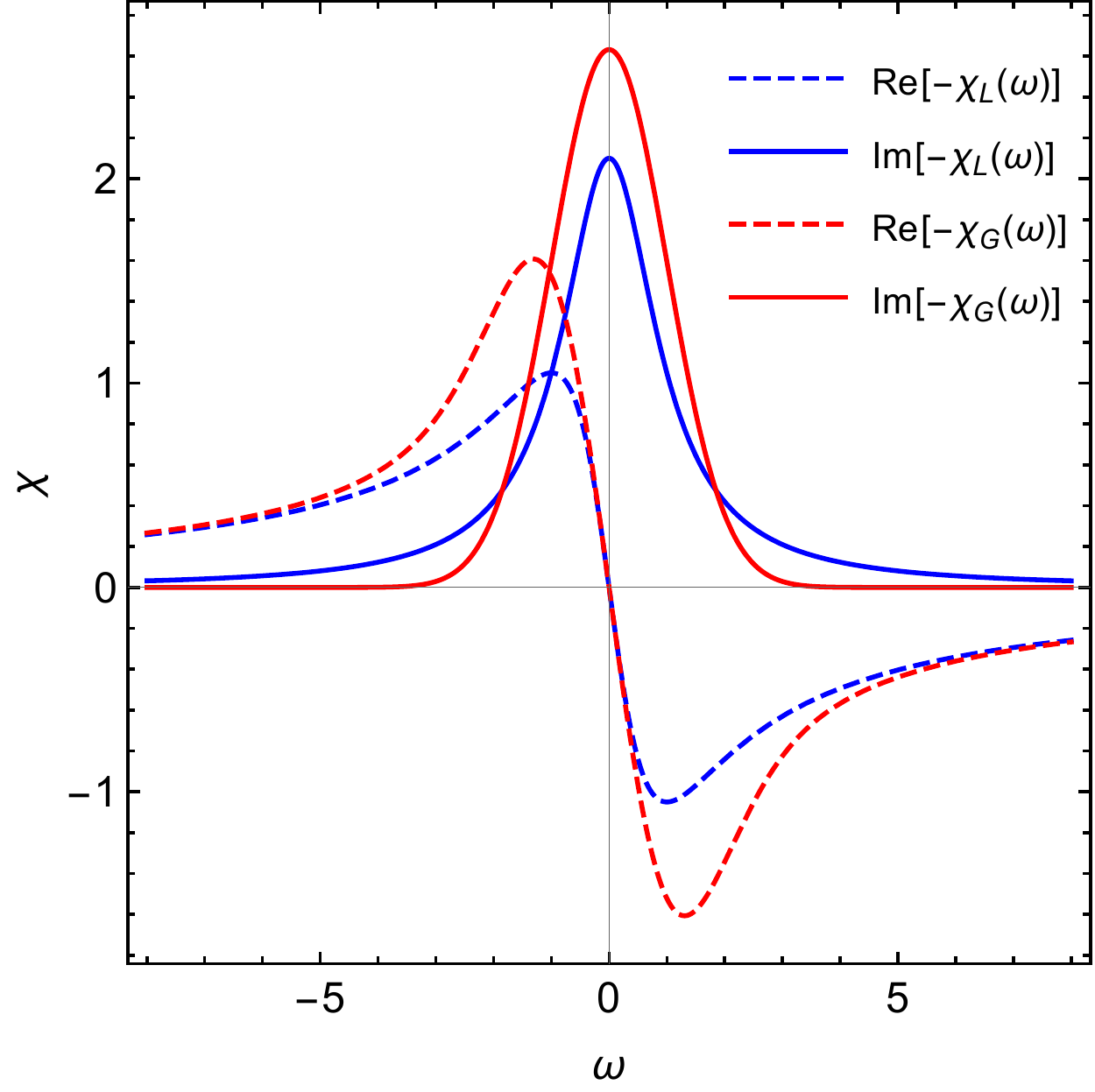}}
\caption{(Color Online) The real and imaginary parts of $\chi(\omega)$ for a Lorentzian $\chi_L(\omega)$ (Eqs. \eqref{lor1} -- \eqref{lor3}) and a Guassian $\chi_G(\omega)$ (Eqs. \eqref{abs1} and \eqref{abs2}) cases with the same strengths $k=1$ and widths $\gamma=1$ of the resonances. The Gaussian absorption decays much faster than the Lorentzian one. }
\label{LorentzianvsGaussian}
\end{figure}

For example, let the absorption be described by a Lorentzian:
  \be
  \chi_2(\omega)= -\frac{k\gamma}{(\omega-\omega_0)^2+\gamma^2}.
  \label{lor1}
  \ee
  where $\omega_0\gg \gamma$ is the optical frequency of the resonance. The minus sign in (\ref{lor1}) is needed to guarantee that the Schr\"odingier equation (\ref{sch1}) leads to a decay of the amplitude.
 Substitituting (\ref{lor1}) into (\ref{kk-1}) we find:
  \be
  \chi_1(\omega)= \frac{k (\omega-\omega_0)}{(\omega-\omega_0)^2+\gamma^2}.
  \label{lor2}
  \ee
 Equations (\ref{lor1}) and (\ref{lor2}) can be written more compactly as
 \be
 \chi(\omega) = \frac{k}{(\omega-\omega_0)+i\gamma}.
 \label{lor3}
 \ee
which makes it clear that the pole of (\ref{lor3}) is at the point $\omega=\omega_0-i\gamma$, i.e. it is placed at the lower complex half-plane.
Generally, we assume that $\chi(\omega)$ can have multiple poles or other singularities in the lower half-plane. For example, consider a Gaussian-broadened shape of absorption:
 \be
 \chi_2(\omega) = -\sqrt\frac{ \pi }{2}\frac{  k}{\gamma} e^{-\frac{(\omega-\omega_0)^2}{2\gamma^2}},
 \label{abs1}
 \ee
 then
 \be
 \chi_1(\omega) =\sqrt\frac{ \pi }{2}\frac{  k}{\gamma}  e^{-\frac{(\omega-\omega_0)^2}{2\gamma^2}}\rm{erfi}\left(\frac{\omega-\omega_0}{\sqrt 2\gamma}\right),\quad  
 \label{abs2}
 \ee
where ${\rm erfi}(x)=\frac{2}{\sqrt \pi}\int_0^y e^{y^2}dy$ is the imaginary error function. For any large $\omega$ in the upper half of the complex plane, including the real axis, we have: 
$\chi_1(\omega) \sim k/(\omega-\omega_0)$ at $|\omega-\omega_0| \gg \gamma$, i.e. it satisfies our physical assumptions.
Such a function $\chi(\omega)$ has no singularities at finite values of $\omega$ but it has exponentially diverging singularity at $\omega \rightarrow \infty$ in the lower complex half-plane. In Fig.~\ref{LorentzianvsGaussian} we compare real and imaginary parts of the Lorentzian and Gaussian response functions. In particular, this figure shows that while imaginary parts have strongly different behavior at large $\omega$, the real parts of $\chi(\omega)$ have similar asymptotics.

The form of $\chi(\omega)$ can possibly be very complex. For example, due to the linearity of Kramers-Kronig relations, any function that is a sum of several above mentioned functions $\chi(\omega)$ taken with different values of constant parameters also satisfies our constraints.

\section{Sweep of the beam frequency}

Imagine now the situation when the frequency of the laser beam is made linearly time-dependent:
\be
\omega-\omega_0 = \beta t.
\label{linear1}
\ee
At $t \rightarrow \pm \infty$, the imaginary part of $\chi(\omega)$ is strongly suppressed, and the real part behaves as $\sim 1/t$, which is still negligible in comparison to the constant coupling to the external magnetic field. Therefore,
if a system is prepared in one of the eigenstates $|1\ra$ or $|2 \ra$ of the Hamiltonian (\ref{ham}) with $\chi(\omega)\rightarrow 0$, transitions between such states would not happen at $t\rightarrow \pm \infty$, and the scattering problem is well defined for a sweep of frequencies throughout the resonance.

Suppose the system is initially (at $t\rightarrow - \infty$) in the eigenstate $|1\ra$ of the Hamiltonian, and we can measure the state of electron in the quantum dot after the sweep, at $t\rightarrow +\infty$. There are 3 possibilities:
First, electron can simply escape from the dot into the continuum. Second, it can stay in the dot and remain in the same state $|1\ra$. Finally, it can stay but flip the spin into the state $|2\ra$.
All those processes have some probabilities to be realized. Let $P_{i\rar j}$ be the probability that the system, starting in the state $|i \ra$, will end up in the state $|j\ra$, where $i,j=1,2$.
Probabilities of all possible outcomes are described by the following 2x2 matrix:
\be
\hat{P}=\left( \begin{array}{cc}
P_{1\rar 1} & P_{2 \rar 1} \\
P_{1 \rar 2} & P_{2 \rar 2}
\end{array}\right).
\label{prm1}
\ee
Note that this matrix is sufficient to determine the probability to escape, e.g.
\be
P_{1 \rar \emptyset} = 1-P_{1 \rar 1}-P_{1 \rar 2}, \quad P_{2 \rar \emptyset} = 1-P_{2 \rar 1}-P_{2 \rar 2},
\label{pr0}
\ee
where $\emptyset$ denotes  the state with electron in the continuum.

Despite the simplicity of the sweep (\ref{linear1}), almost arbitrary complexity of the function $\chi(\omega(t))$ makes the problem to determine the scattering probability matrix (\ref{prm1}) look extremely complex. For example, we know that one cannot  solve a problem of a 2-state system driven by arbitrary time-dependent fields analytically and exactly. Nevertheless, in this article we show that no matter how complex the function $\chi(\omega)$ can be, as far as it satisfies the above mentioned constraints, we  can provide exact formulas for three out of four elements of the matrix (\ref{prm1}).

\section{Solution of the model}

\begin{figure}
\scalebox{0.12}[0.12]{\includegraphics{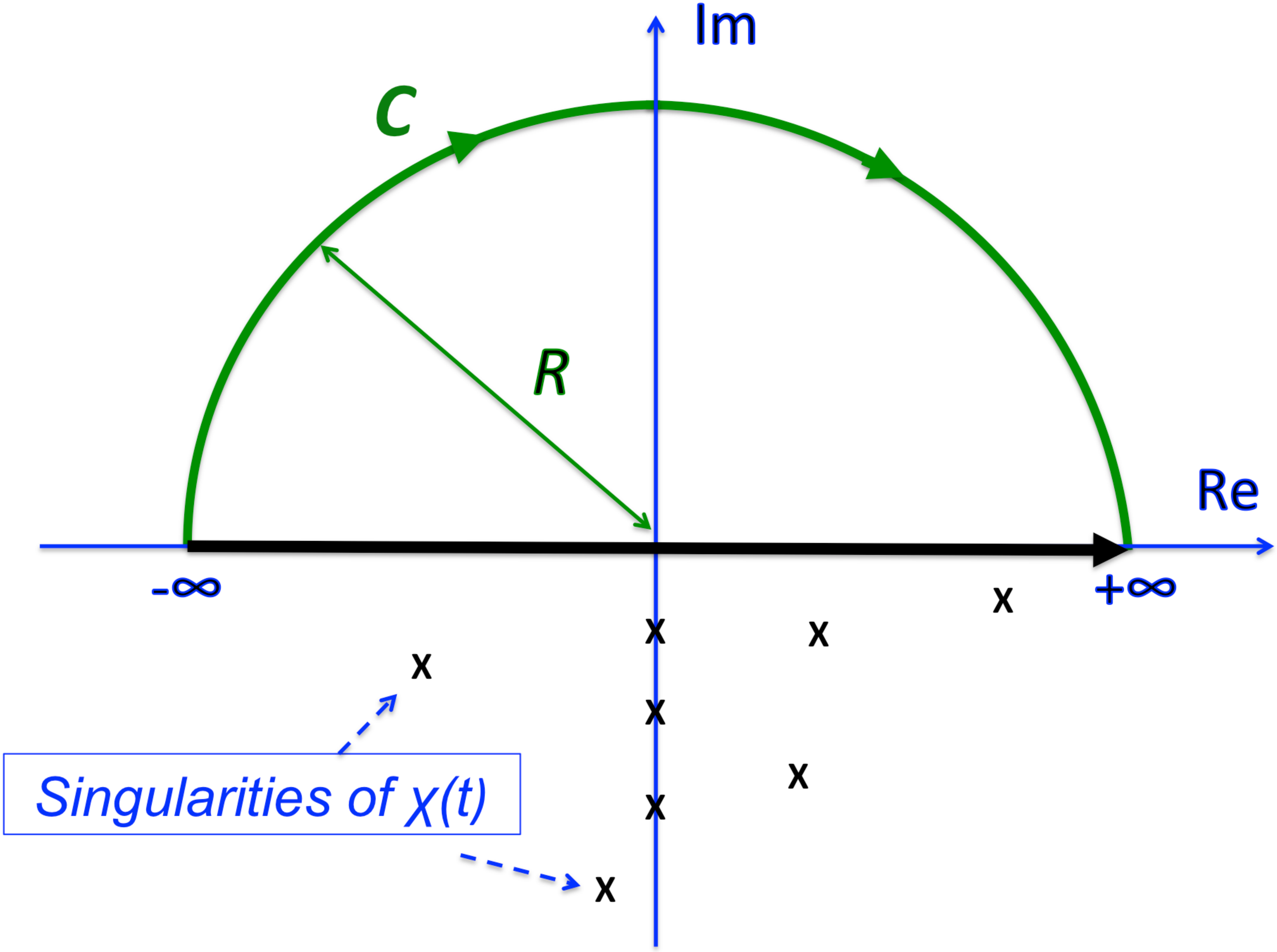}}
\hspace{-2mm}\vspace{-4mm}
\caption{ (Color online) The time evolution contour along the real axis (black) can be continuously deformed into a semicircle  with radius $R\rightarrow \infty$ laying in the upper complex half-plane (green) without affecting initial conditions at real values of $t\rightarrow -\infty$. Since the function $\chi(t)$ is analytic in the upper-half plane,  this deformation does not encounter the singularities of the Hamiltoninan so that the scattering matrix does not change during the time contour deformations. Analogous continuation to the lower half-plane would be impossible because of singularities of $\chi(t)$.}
\label{contour-fig}
\end{figure}
We now solve the model using a contour deformation method described in \cite{sinitsyn-14pra}. This method takes roots in the Landau's derivation of the Landau-Zener (LZ) formula \cite{landau}. First, since we are looking at transitions between eigenstates of the magnetic field, it is better to switch to the basis of these states. Let $\theta$ be the angle between the magnetic field and the beam axis. Then the Hamiltonian reads:
\be
\hat{H} = \frac{\varepsilon}{2} \hat{\sigma}_z +\frac{\chi(\omega (t))}{2} \left( \hat{1}+ \cos (\theta) \hat{\sigma}_{z} +\sin (\theta) \hat{\sigma}_x \right),
\label{ham2}
\ee
where $\varepsilon$ is the parameter that describes the energy splitting by the external magnetic field and $\hat{1}$ is the unit 2x2 matrix; $\omega(t)$ is given by (\ref{linear1}). Without a loss of generality, we will set $\varepsilon>0$.

Since the Hamiltonian (\ref{ham2}) has no singularities in the upper plane, the solution of the Schr\"odinger equation is analytic in this region. Hence we can deform the time contour continuously so that it has the form of a semicircle ${\bf C}$ with radius $R\gg \omega_0,\gamma,\varepsilon$, as shown in Fig.~\ref{contour-fig}.  This contour connects the same initial and final points at real values of $t\rightarrow \pm \infty$, so the solution along this contour can be used to solve the desired scattering problem.

Along the contour ${\bf C}$, the function $\chi(\omega)$ behaves as
\be
\chi(\omega)|_{\bf C} =\frac{k}{\beta t} +o(1/R) ,
\label{chi2}
\ee
where $k$ is the real constant defined previously.

Let, initially, the system be in the state $|1\ra$ with energy $\varepsilon/2$. Following \cite{sinitsyn-14pra}, we can say that since the off-diagonal elements of the Hamiltonian along the contour ${\bf C}$ become negligibly small, we can apply adiabatic approximation everywhere along the contour ${\bf C}$, which corresponds in our case to   keeping only contributions to the adiabatic energy that decay not faster than $\sim 1/R$ at $R\rightarrow \infty$.
In particular, this means that we can disregard the off-diagonal terms of the Hamiltonian.
The adiabatic approximation leads then to the solution for the amplitude to remain in the state $|1 \ra$ along the contour ${\bf C}$:
\be
\psi_1(\tau) = \exp \left( -i \int_{-\infty +i0}^{\tau}  \Big[ \frac{\varepsilon}{2} +\frac{ k(1+ \cos (\theta))}{\beta t} \Big]\, dt \right).
\label{sol1}
\ee
An analogous  solution is obtained for the amplitude to remain in the state $|2\ra$ if the spin system is prepared in this state at $t\rightarrow -\infty$:
\be
\psi_2(\tau) = \exp \left( -i \int_{-\infty +i0}^{\tau}  \Big[ -\frac{\varepsilon}{2} +\frac{k( 1- \cos (\theta))}{\beta t} \Big]\, dt \right).
\label{sol2}
\ee
We note that although the form of $\chi(\omega(t))$ can be very complicated, Eq.~(\ref{sol1}) is valid as long as Eq.~\eqref{chi2} is satisfied, since the arguments leading to that equation only depend on asymptotic behavior of the Hamiltonian at $R\rar \infty$.

\begin{figure}
\scalebox{0.5}[0.5]{\includegraphics{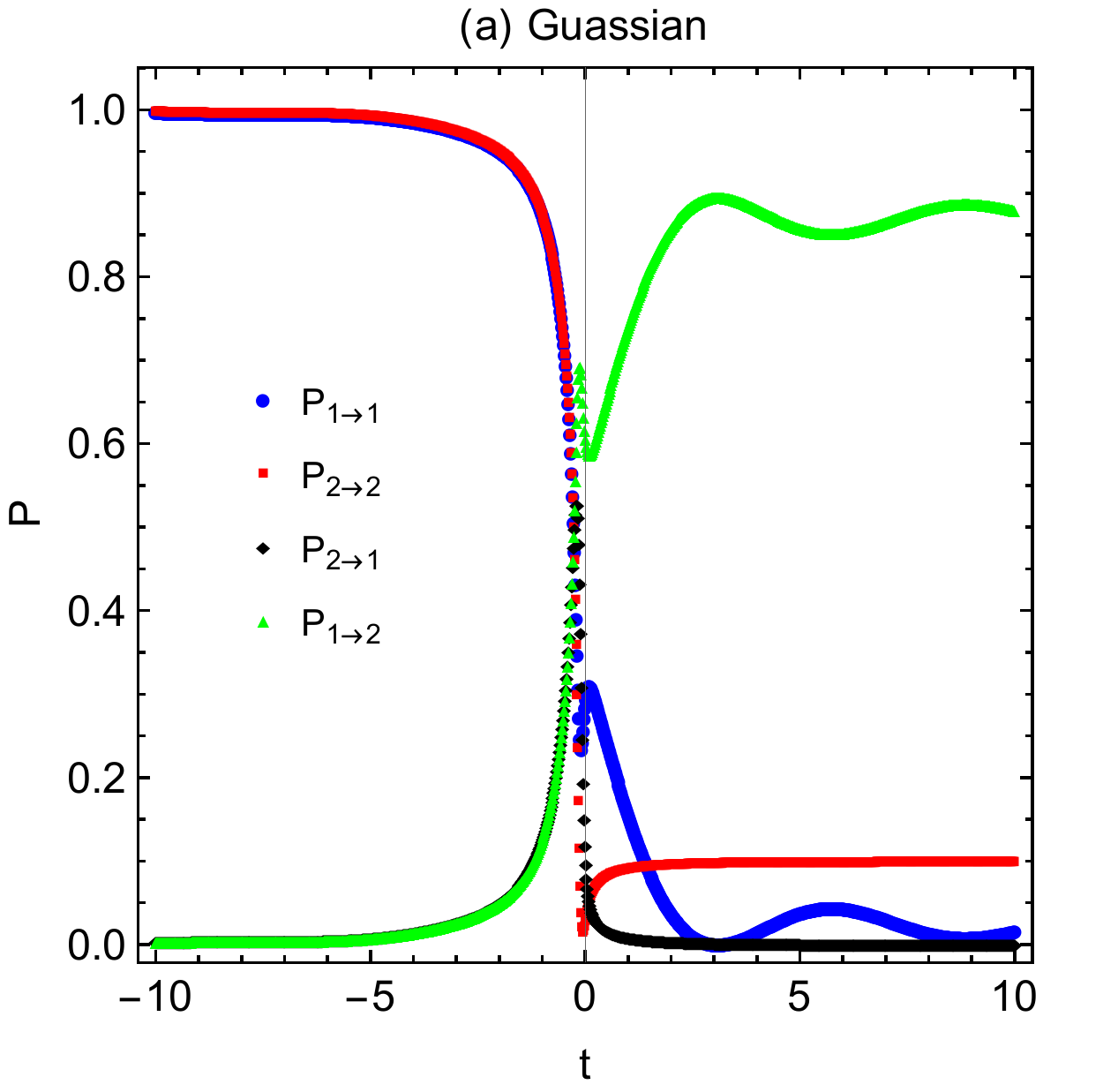}}
\scalebox{0.5}[0.5]{\includegraphics{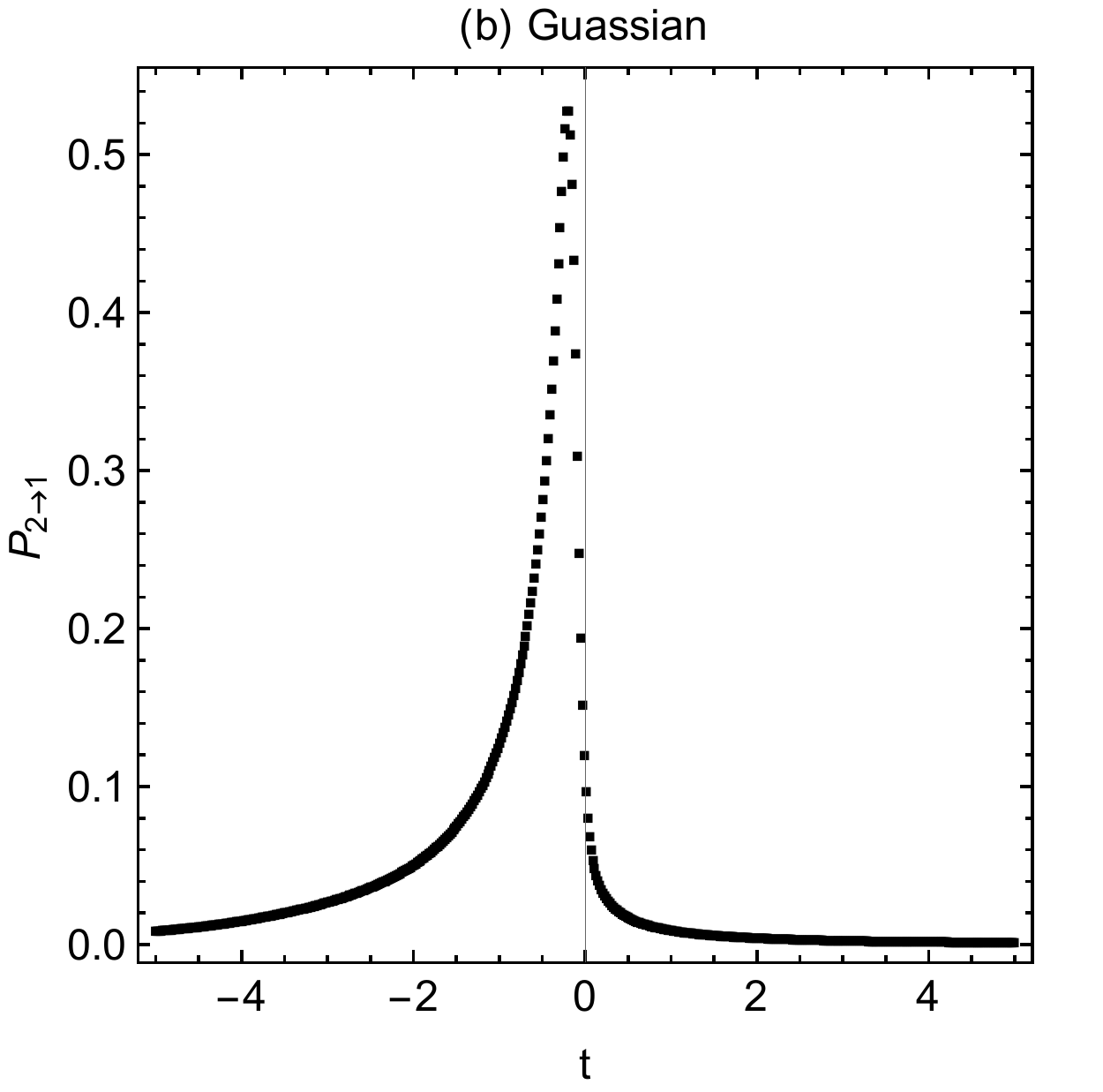}}\\
\scalebox{0.5}[0.5]{\includegraphics{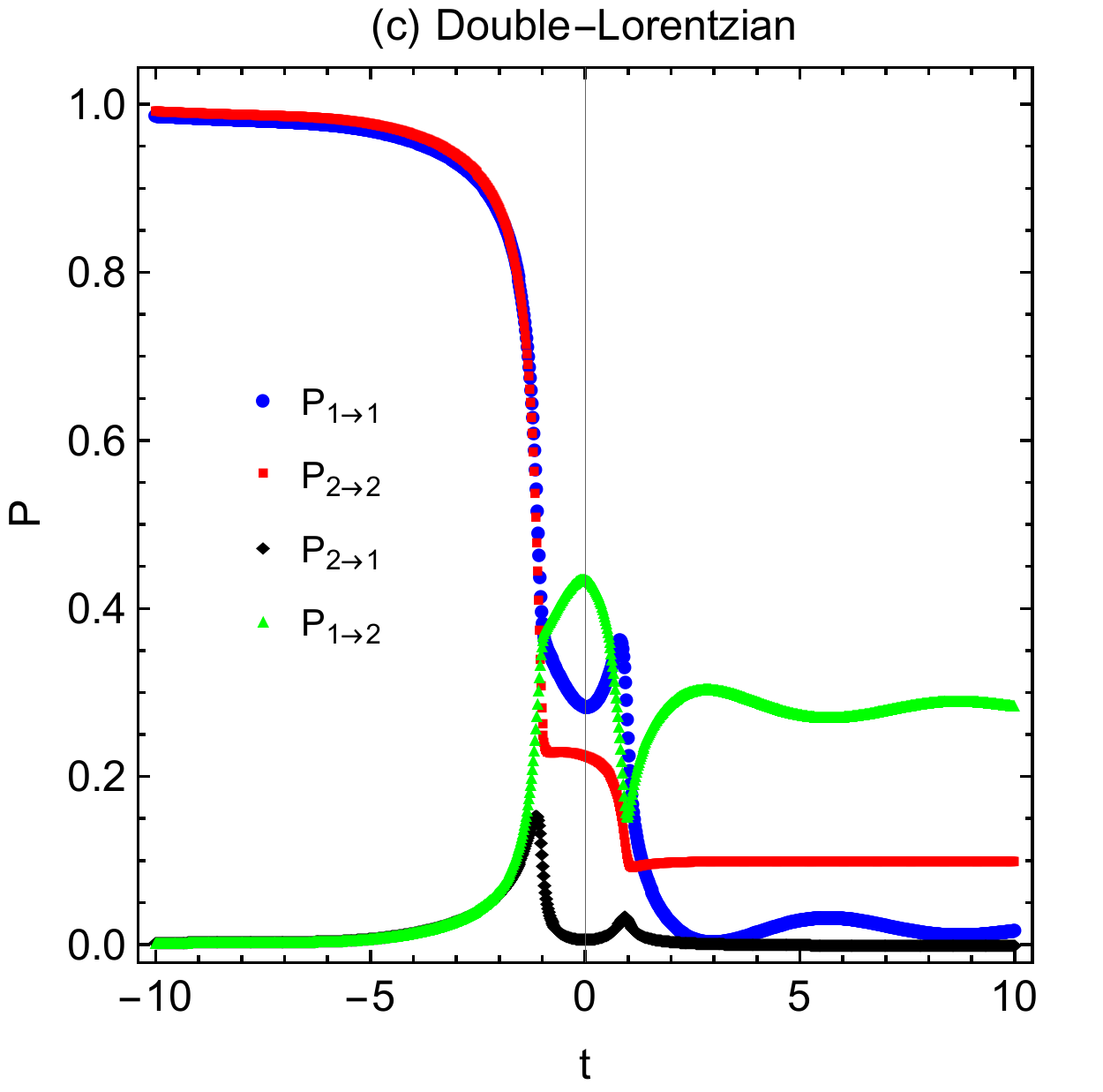}}
\scalebox{0.515}[0.515]{\includegraphics{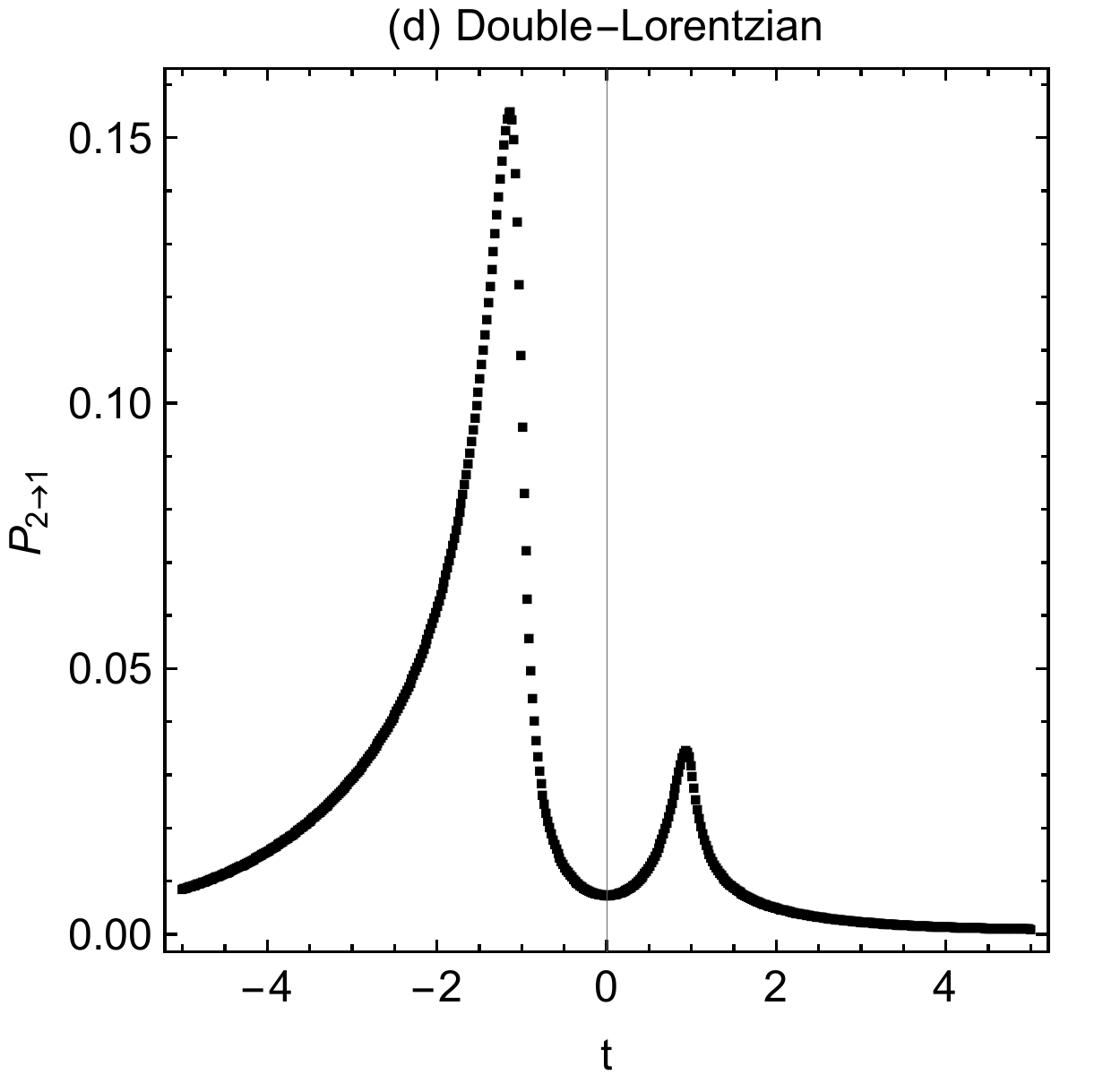}}\\
\scalebox{0.505}[0.505]{\includegraphics{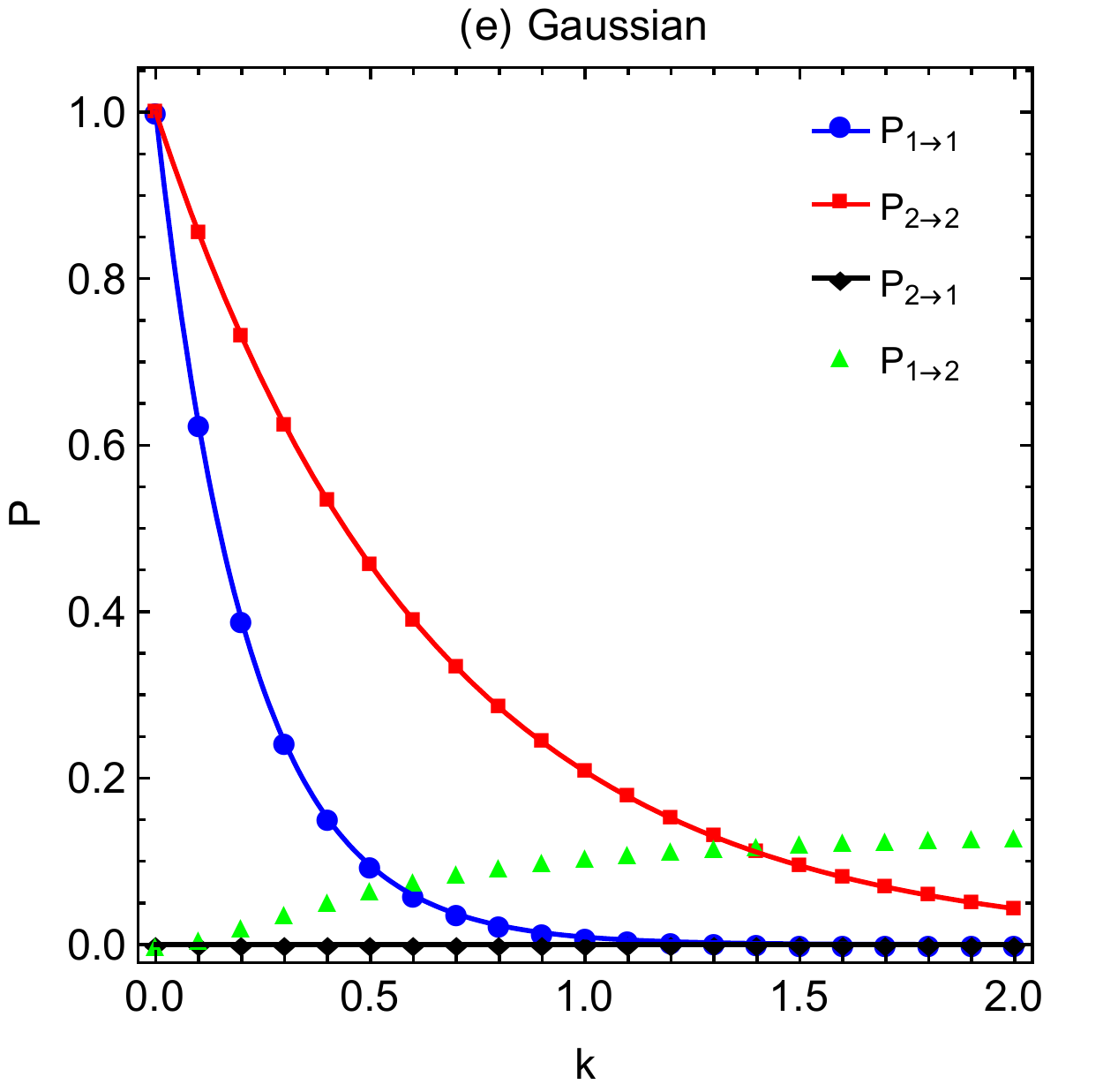}}
\scalebox{0.505}[0.505]{\includegraphics{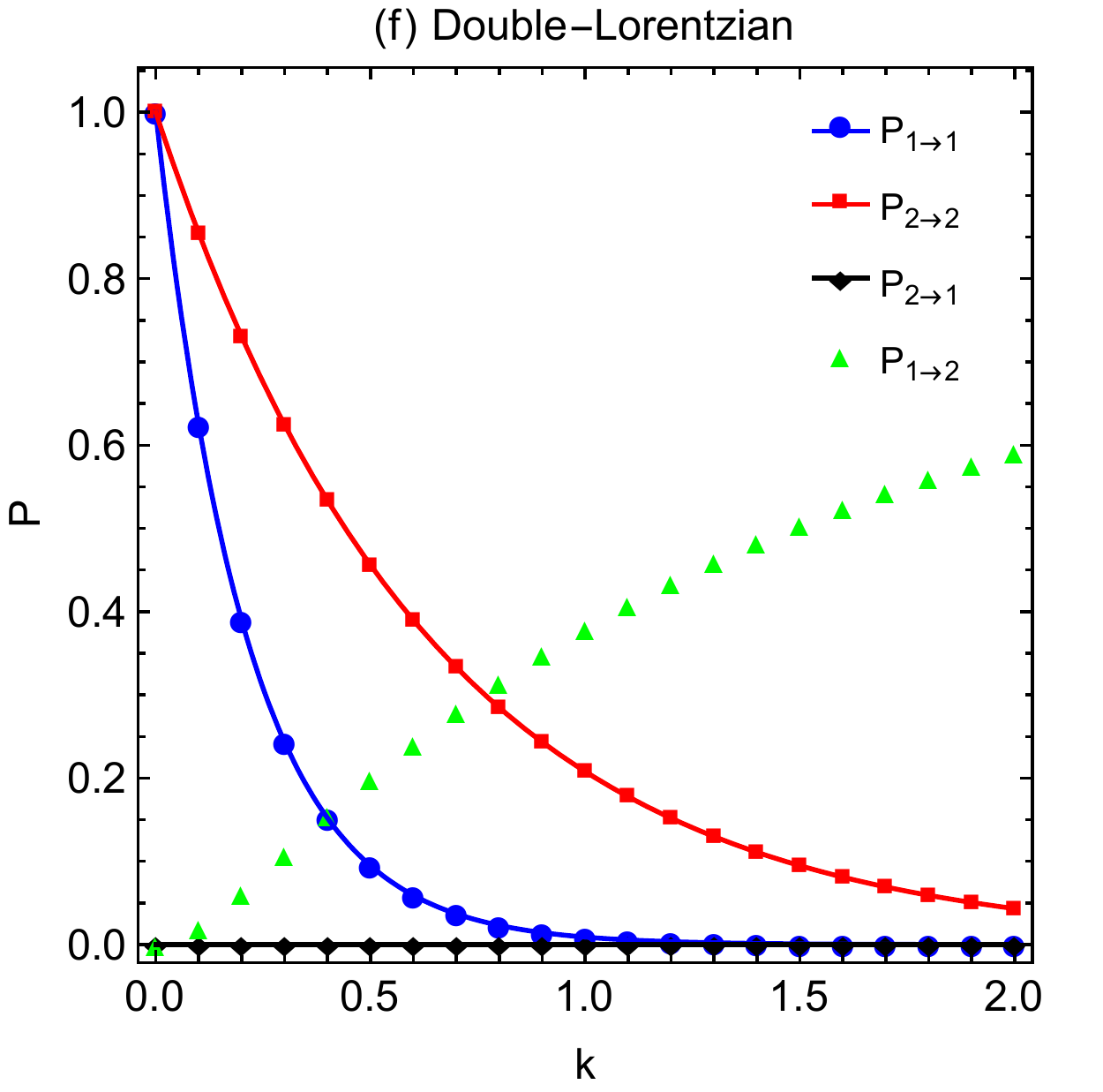}}
\caption{(Color Online)  Numerical tests of analytical results. The dots correspond to direct numerical solutions of the Schr\"odinger equation (\ref{sch1}), and the solid curves correspond to theoretical predictions in Eq.~(\ref{prm2}). In all cases, we make the choice of parameters $\beta=\epsilon=1$. (a) Time dependence of the four transition probabilities for the Guassian case. Choice of parameters: $k=1$, $\gamma=0.1$, $\theta=5\pi/12$. The time-interval for simulations is from $t=-50$ to 50 and the time step is 0.0002. Only the curves in the interval from $t=-10$ to $10$ are shown. Each probability undergoes a sharp change at $t=0$ and then oscillates before saturation. (b)  Time dependence of the probability $P_{2\rar1}$ zoomed in near $t=0$. Note that although it approaches zero asymptotically at $t\rar +\infty$, it is nonzero at intermediate time moments. (c) Same as (a) but for the ``Double-Lorentzian'' case: $\chi(\omega)=\frac{1}{2}(\frac{k}{\omega-\omega_1+i\gamma}+\frac{k}{\omega-\omega_2+i\gamma})$, where we choose $\omega_1=-\omega_2=1$. (d) Same as (b) but for the Double-Lorentzian case. Note the existence of two peaks. (e) Transition probabilities
as functions of the strength k for the Gaussian case. Choice of parameters: $\gamma=1$, $\theta=\pi/3$. The time-interval for simulations is from $t=-500$ to 500 and the time step is 0.002. The numerics show perfect agreement with theoretical predictions. (f)  same as (e) for the Double-Lorentzian case.
}
\label{numerical}
\end{figure}

As for the amplitude of the transition from one spin state to the  other state, trimming off-diagonal elements of the Hamiltonian for the evolution along the contour ${\bf C}$ leads to the prediction of a zero transition probability between states $|1\ra$ and $|2\ra$. However, according to Ref.~\cite{sinitsyn-14pra}, this expectation is valid only for the transition from the state $|2\ra$, which has the lower adiabatic energy along the contour ${\bf C}$, to the state $|1\ra$. In terminology of \cite{sinitsyn-14pra}, the transition $|2\ra \rar |1\ra$ is ``counterintuitive''. The discussion of exact suppression of counterintuitive transitions and the reasons why the adiabatic approximation fails
 in our case for the transition from $|1\ra$ into $|2\ra$ would be completely analogous to the discussion of the no-go theorem in \cite{sinitsyn-14pra,no-go}.

By performing integration in (\ref{sol1})-(\ref{sol2}) along the full contour ${\bf C}$, and taking into account the no-go theorem, we  obtain the following scattering matrix elements:
\be
S_{11} = e^{-\pi k(1+\cos (\theta))/(2\beta)}, \quad S_{12}=0, \quad S_{22}=e^{-\pi k (1-\cos(\theta))/(2\beta)}.
\label{sc1}
\ee
Taking the absolute value squared of the scattering matrix elements, we find the form of the  transition probability matrix:
\be
\hat{P}=\left( \begin{array}{cc}
 e^{-\pi k(1+\cos (\theta))/\beta}& 0 \\
{\rm X} & e^{-\pi k(1-\cos (\theta))/\beta}
\end{array}\right),
\label{prm2}
\ee
where X indicates that this element of the transition probability matrix cannot be determined by the above described procedure.

Equation~\eqref{prm2} is the main result of this article. It shows that three out of four independent transition probabilities are described by simple formulas. In Fig.~\ref{numerical}(a-d) we show numerical results for time dependence of transition probabilities
in the models with Gaussian and ``Double-Lorentzian". In particular, they demonstrate that at intermediate time moments the transition from the state $|2\ra$ into the state $|1\ra$ can be nonzero but it approaches zero value asymptotically at $t\rar +\infty$, i.e. the zero value of the corresponding matrix element in (\ref{prm2}) is not trivial. In Fig.~\ref{numerical}(e) and (f) we compare predictions of Eq.~(\ref{prm2}) with results of our direct numerical simulations.

\begin{figure}
\scalebox{0.505}[0.505]{\includegraphics{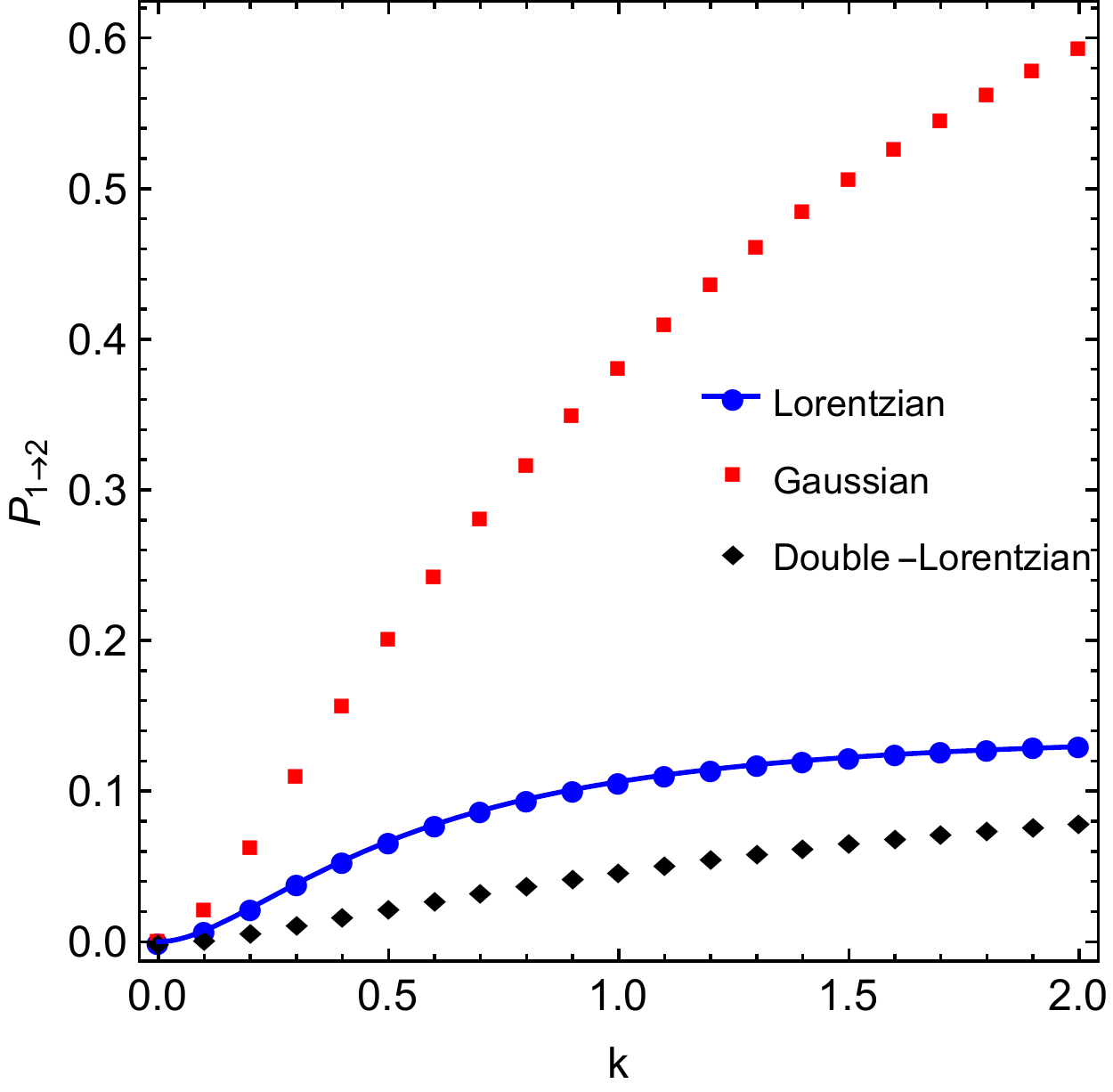}}
\caption{(Color Online) Numerical results (the dots) for the probabilities $P_{1\rar2}$ (or X) for three different cases with the same strength $k=1$: Lorentzian, Gaussian and ``Double-Lorentzian''. The choice of other parameters is $\beta=\gamma=\epsilon=1$, $\theta=\frac{\pi}{3}$, and $\omega_1=-\omega_2=1$ for double-Lorentzian. The time-interval for simulations is from $t=-500$ to 500 and the time step is 0.002. For the Lorentzian, solid curve represents the theoretical prediction of Eq.~\eqref{lorX}. }
\label{nonuniversal}
\end{figure}

It is likely that a simple exact analytical expression for the remaining unknown element of the transition probability matrix (\ref{prm2}) does not exist. In order to shed light on its behavior, we identified one fully solvable
model with the Hamiltonian of the type (\ref{ham2}), which complete solution we worked out in Appendix B. The model corresponds to the Lorentzian shape of the dissipative part of the function $\chi(\omega)$ (Eqs. \eqref{lor1} -- \eqref{lor3}). The corresponding Schr\"odinger equation can be mapped to the confluent hypergeometric equation, which solution has well-understood asymptotics. The result for the non-universal component then reads:
\begin{align}\label{lorX}
{\rm X}=e^{-2\varepsilon\gamma/\beta}[1-e^{-\pi k(1+\cos\theta)/\beta}][1-e^{-\pi k(1-\cos\theta)/\beta}].
\end{align}
Equation~\eqref{lorX} explicitly shows that X depends on both $\varepsilon$ and $\gamma$. Figure~\ref{nonuniversal} shows that the matrix element denoted by X is not universal. It depends sensitively on all parameters of the model, in particular on the choice of the functional form of $\chi(\omega)$. Nevertheless, other scattering probabilities are independent of $\varepsilon$ (as far as it is positive), as well as numerous details that can describe the function $\chi(\omega)$.


\section{Extension to multistate systems}

\begin{figure}
\scalebox{0.5}[0.5]{\includegraphics{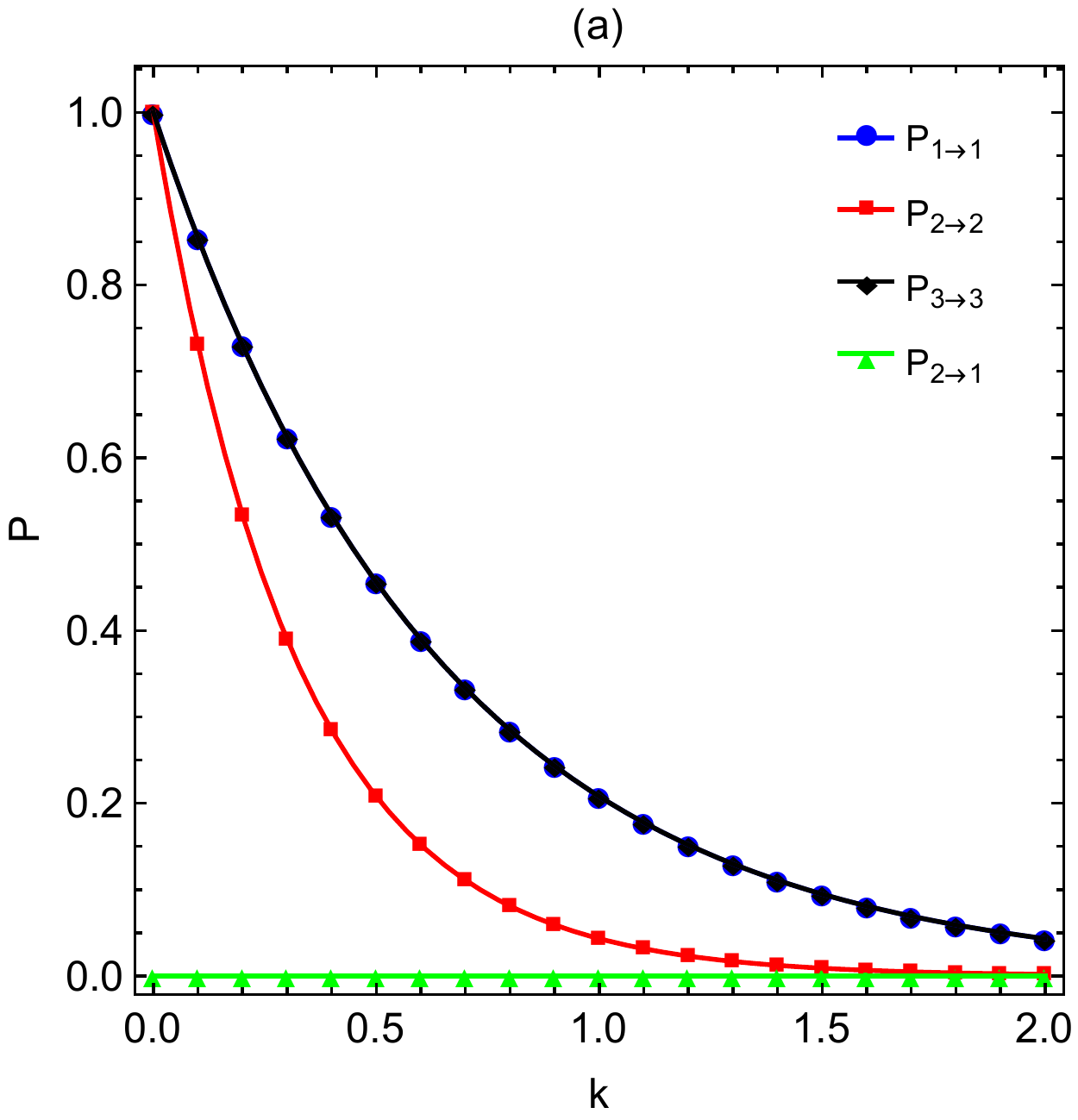}}
\scalebox{0.51}[0.51]{\includegraphics{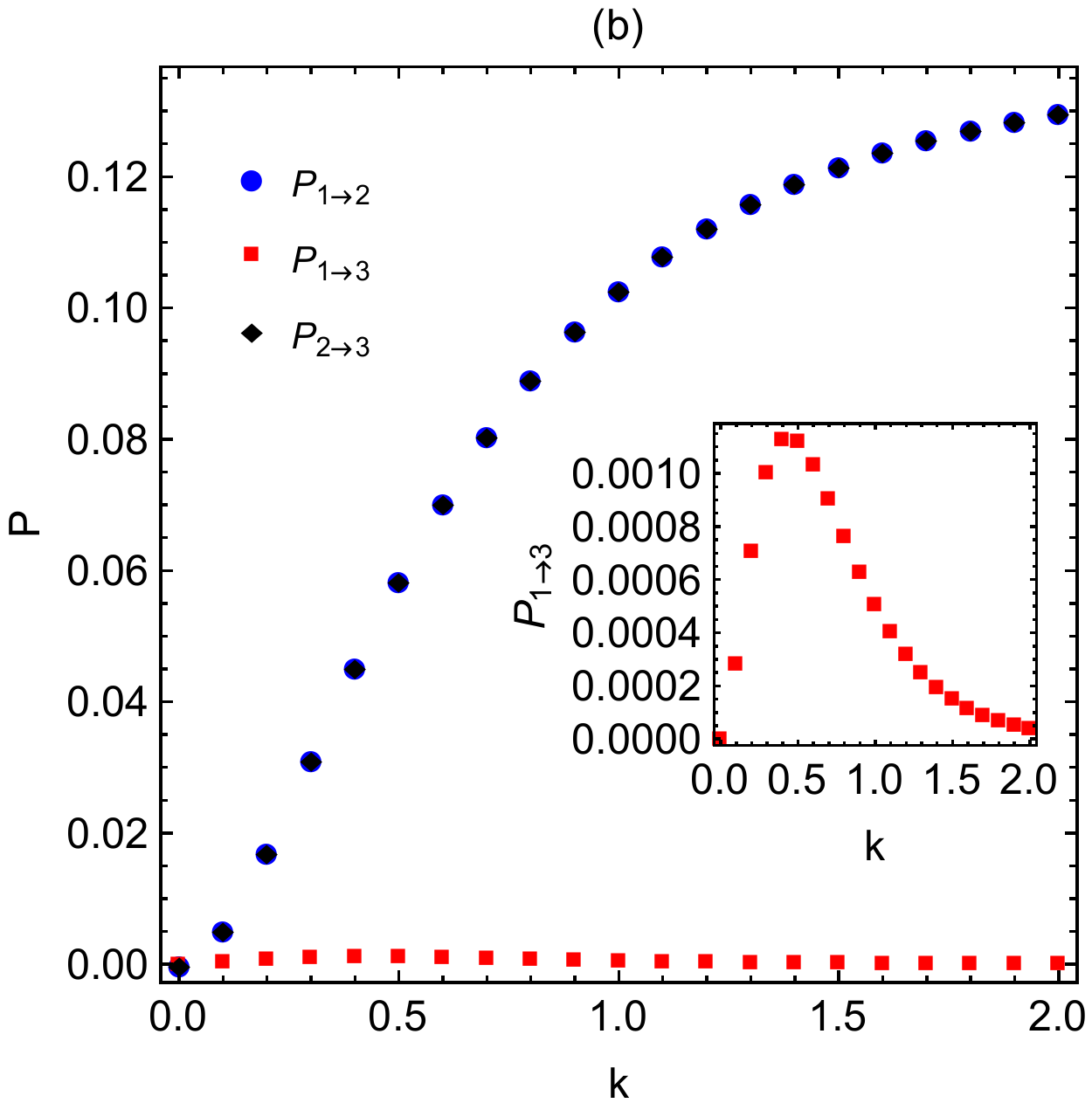}}\\
\caption{(Color Online)  Numerical results for a three-level system with Hamiltonian Eq. \eqref{ham32}, where $\chi$ is taken to be Lorentzian. The dots correspond to direct numerical solutions and the solid curves correspond to theoretical predictions in Eq.~(\ref{prm3}). Choice of parameters: $\beta=\gamma=\epsilon=1$. The time-interval for simulations is from $t=-500$ to 500 and the time step is 0.002. (a) Analytically determined transition probabilities as functions of the strength $k$. Here $P_{1\rar1}$ and $P_{3\rar 3}$ coincide. $P_{3\rar 1}$ and $P_{3\rar 2}$ are not shown but both of them are tested to be zero. (b) The three non-universal transition probabilities as functions of $k$. $P_{1\rar2}$ and $P_{2\rar 3}$ coincide. The inset shows $P_{1\rar3}$ which is significantly smaller than the other two probabilities.
}
\label{3level}
\end{figure}

Our results are straightforward to generalize to multistate systems linearly driven through a resonance \cite{do,be}. For example, consider a three-level system (possibly a spin-1 multiplet of an electron in a quantum dot molecule) with a Hamiltonian
\begin{equation}
\hat{H} = \left( \begin{array}{ccc}
\chi(\omega) & 0 & 0\\
0& 0 &0\\
0& 0 &0
\end{array}\right) + \varepsilon\hat{S_x}.
\label{ham3l}
\end{equation}
For simplicity, we assumed here that the light interacts with only one of the states. In the basis of eigenstates of the Hamiltonian with $\chi(\omega)=0$ the Hamiltonian matrix   reads:
\begin{equation}
\hat{H} = \left( \begin{array}{ccc}
\varepsilon & 0 & 0\\
0 & 0 &0\\
0 & 0 & -\varepsilon\\
\end{array}\right)
\label{ham32}+
\frac{\chi(\omega)}{4}\left( \begin{array}{ccc}
1 & -\sqrt 2 & 1\\
-\sqrt 2 & 2 &-\sqrt 2\\
1 & -\sqrt 2 & 1\\
\end{array}\right).
\end{equation}
Repeating the same arguments as in the two-level case, and using the no-go theorem from \cite{sinitsyn-14pra}, we find the form of the transition probability matrix:
\be
\hat{P}=\left( \begin{array}{ccc}
 e^{-\pi k/(2\beta)}& 0 & 0\\
{\rm X_{1\rar2}} &  e^{-\pi k/\beta} &0\\
{\rm X_{1\rar3}} & {\rm X_{2\rar3}} & e^{-\pi k/(2\beta)}
\end{array}\right).
\label{prm3}
\ee
There are three unknown elements in (\ref{prm3}) but all other transition probabilities are determined exactly regardless of the details of $\chi(\omega)$. The numerical check for this model, in support of Eq. \eqref{prm3}, is presented in Fig.~\ref{3level}(a).  Figure~\ref{3level}(b) shows numerical results of the three unknown transition probabilities. Equation~\eqref{prm3} can be readily generalized to an arbitrary n-level system interacting  via a resonant transition.

We would like to note that there are potential applications of
this multistate generalization to the field of ultracold atomic gases interacting via the Feshbach resonance. The latter corresponds to a coherent creation of molecules from pairs of single atoms. The strength of atomic interactions via such a resonance can be tuned by a magnetic field and has a typical $\sim 1/(B-B_0)$ dependence on the external field. There already have been numerous experiments performed to study the effect of a linear sweep of the magnetic field through the Feshbach resonance in order to create a molecular condensate \cite{feshbach}. Certainly, direct application of our results to such systems will need additional justification because creation of a molecule may not be possible to describe as a dissipative transition. However, phenomenologically, if we identify the discrete states with quantum states of an atomic gas, our results predict, in particular, that the sweep through the Feshbach resonance cannot leave the  atomic part of the gas in an excited state. Moreover, we can obtain the formula for the probability to remain in the initial state after the sweep through the resonance. If the atomic gas is initially in the ground state, then combining such predictions   we can derive the probability of not creating any molecule after the sweep through the Feshbach resonance e.t.c.. In fact, the possibility of such results is in agreement with some of the existing theoretical publications on this topic \cite{dobrescu,gurarie,chain}.

\section{Conclusion}

We considered a localized electronic spin controlled by a circularly polarized beam and an external magnetic field. We assumed that the frequency of the beam is made linearly time-dependent so that it sweeps through the optical resonance, starting and ending at the values far away from it. Causality of the spin response to the optical interactions imposes the Kramers-Kronig relations on the response function and make it possible for the electron to escape into a continuum of states. Using a contour deformation method, we found exact probabilities of transitions between different possible electronic states after such a sweep of the optical frequency. These probabilities do not depend on the shape of the resonance but rather they depend on its integrated power $k$ of the resonance and the relative angle between the optical beam and the applied external magnetic field. This simplicity can be used in practice, e.g. to estimate the parameter $k$ experimentally. We showed that this result can be generalized to multistate systems which can be useful, e.g., for studies of transitions through the Feshbach resonance in ultra cold atomic gases.
We believe that our theoretical predictions can be verified experimentally in spin systems controlled by optical beams, e.g.,  in NV centers \cite{awschalom} or in electron doped InGaAs self-assembled quantum dots \cite{alex-nat}.

{\it Acknowledgment}. The work
was carried out under the auspices of the National Nuclear
Security Administration of the U.S. Department of Energy at Los
Alamos National Laboratory under Contract No. DE-AC52-06NA25396. Authors also thank the support from the LDRD program at LANL.

 \section*{Appendix A: localized electron coupled to a continuum of conduction states}
   As a physical example, consider an electron state $|1/2 \ra $ localized in a quantum dot of a semiconductor interacting with a circularly polarized beam. This light couples the state $|1/2 \ra $ to the trion exciton with the total spin 3/2. We assume that this excited state is not stable so that if it is created, one electron quickly escapes into the conduction band leaving zero charge in the quantum dot, which makes this quantum dot optically inactive during the time of experiment. Therefore, it is convenient to write the exciton state in the basis of a continuum of states of conduction electrons, $|i\ra$, where index $i$ runs throughout all Bloch states. The Hamiltonian of interaction of the localized electron and the continuum can be written then in the form:
 \begin{equation}
 \hat{H}_{tot}=\omega_{1/2} |1/2\ra \la 1/2| + \sum_i E_i |i\ra\la i | +\left(e^{i\omega t} \sum_i g_i |i\ra\la 1/2 | +h.c. \right),
 \label{hcont}
 \end{equation}
 where $g_i$ are the matrix elements of the dipole operator between localized electron and conduction band states. Such couplings are also proportional to the beam field amplitude. Here we set the energy of the state $|-1/2\ra$ to zero. Since it does not interact with the given circalar polarization, this state drops out from (\ref{hcont}). As shown in Fig.~(\ref{setup-fig}), we assume that the gap between states with energies $E_i$ and $\omega_{1/2}$ is in the optical spectrum so that $\omega \sim E_i-\omega_{1/2} \gg |E_j- E_i|$.  We will also assume that the couplings $g_i$ are weak in comparison to the broadening of the exciton state.

Our goal is to derive the effective Hamiltonian describing the localized spin. For this, first, we  redefine the phase of the band states: $|i\ra \rightarrow e^{-i\omega t} |i\ra$. This makes the Hamiltonian time independent:
 \begin{equation}
 \hat{H}_{tot}=\omega_{1/2} |1/2\ra \la 1/2| + \sum_i (E_i-\omega) |i\ra\la i | +\left(\sum_i g_i |i\ra\la 1/2 | +h.c. \right).
 \label{hcont2}
 \end{equation}
We  use then the stationary perturbation theory to derive the renormalization of the energy of the spin state $|1/2\ra$:
\begin{equation}
\omega_{1/2} \rightarrow \omega_{1/2} - \mP \int dE \rho(E) \frac{|g(E)|^2}{E-\omega -\omega_{1/2}},
\label{renorm}
\end{equation}
where we switched from the sum over discrete index $i$ to the integral over energies $E$ of delocalized states, and introduced the density of states $\rho(E)$ and the energy-dependent coupling strength $g(E)$.

Equation~(\ref{renorm}) includes only virtual transitions to the exited states. However, there are inevitable dissipative transitions, which rate is given by the golden rule:
\begin{equation}
\frac{d|\psi_{1/2}|^2}{dt} =- 2\pi \int dE \, \rho(E) |g(E)|^2 \delta \left(E-\omega - \omega_{1/2} \right),
\label{golden}
\end{equation}
and which can be introduced in the Schr\"odinger equation by adding a complex energy to the state $|1/2 \ra$:
\begin{equation}
\omega_{1/2} \rightarrow \omega_{1/2} -i\pi \int dE \, \rho(E) |g(E)|^2 \delta \left(E-\omega - \omega_{1/2} \right).
\label{golden2}
\end{equation}
Combining Eqs.~(\ref{renorm}) and (\ref{golden2}) we find $\omega_{1/2} \rightarrow \omega_{1/2} +\chi(\omega)$, where
\begin{equation}
\chi(\omega)=-\mP \int dE \rho(E) \frac{|g(E)|^2}{E-\omega -\omega_{1/2}} -i\pi \int dE \, \rho(E) |g(E)|^2  \delta \left(E-\omega - \omega_{1/2} \right),
\label{chi-spin}
\end{equation}
where the first and the second terms in (\ref{chi-spin}) can be identified with, respectively, the dissipationless $\chi_1(\omega)$ and dissipative $\chi_2(\omega)$ parts of the response function $\chi(\omega)$.
To show that they satisfy the Kramers-Kronig relations, it is sufficient to consider a special case $\rho(E) =\delta(E-E_0)$  (because Kramers-Kronig relations are linear, and an arbitrary density $\rho(E)$ can be represented as a continuum of such delta-functions).
In this case
\begin{equation}
\chi(\omega)= |g(E_0)|^2 \Big[- \frac{1}{E_0-\omega -\omega_{1/2}} -i\pi  \delta \left(E_0-\omega - \omega_{1/2} \right) \Big].
\label{chi-spin}
\end{equation}
 It is then straightforward to verify that relations (\ref{kk-2}) hold for corresponding $\chi_1(\omega)$ and $\chi_2(\omega)$.
One can also rewrite
$$
\chi(\omega)=\frac{|g(E_0)|^2}{\omega -E_0+\omega_{1/2}+i \delta}, \quad \delta \rightarrow 0^+,
$$
so that $\chi(\omega)$ clearly has only a single pole in the lower half of the complex plane.

\section*{Appendix B: Transition probabilities for the Lorentzian shape of the resonance}
Consider the Hamiltonian \eqref{ham2} with a Lorentzian shape of $\chi_2(\omega)$. Under the linear sweep of the frequency,
$$
\chi(t) =k/(\beta t +i\gamma).
$$
We will assume that $\varepsilon>0$. Let us write the state vector as $\psi=(\varphi_1,\varphi_2)^T$. The change of variables $t \rar t - i\gamma/\beta$ eliminates the parameter $\gamma$ from the evolution equation (but moves it to the boundary conditions). Eliminating $\varphi_1$ from the Schr\"odinger equation \eqref{sch1}, we get a 2nd order differential equation for $\varphi_2$:
\begin{align}\label{eqphi2}
&t\varphi_2''+\left(1+i\frac{k}{\beta}\right)\varphi_2'+\left(-\frac{i}{2}\varepsilon+ \frac{\varepsilon k\cos\theta}{2\beta}+ \frac{\varepsilon^2}{4}t \right)\varphi_2=0.
\end{align}
This equation can be transformed to the  confluent hypergeometric equation:
\begin{align}\label{}
&zc''(z)+(r-z)c'(z)-sc(z)=0,
\end{align}
where $z=-i\varepsilon t$, $c(t)=e^{-i\frac{\varepsilon}{2}t}\varphi_2(t)$, $r=1+i\frac{k}{\beta}$ and $s=i\frac{k(1-\cos\theta)}{2\beta}$, which has a solution
\begin{align}\label{}
&\varphi_2(t)=e^{-\frac{\tau}{2}}[A M(s,r,\tau)+B U(s,r,\tau)],
\end{align}
where $\tau=-i\varepsilon t+\frac{\varepsilon\gamma}{\beta}$, $M(s,r,\tau)$ and $U(s,r,\tau)$ are the Kummer's and Tricomi's functions, respectively, and $A$ and $B$ are constants to be fixed by initial conditions. The solution for the other state amplitude reads:
\begin{align}\label{}
&\varphi_1(t)=-\tan\frac{\theta}{2} e^{-\frac{\tau}{2}}\{A[M(s,r,\tau)+\frac{\tau}{r} M(s+1,r+1,\tau)]+B[U(s,r,\tau)-\tau U(s+1,r+1,\tau)]\}.
\end{align}

The asymptotic behaviors of the functions $M(s,r,\tau)$ and $U(s,r,\tau)$ at large $|\tau|$ read:
\begin{align}\label{confluentasymp}
&M(s,r,\tau)\sim\frac{\Gamma(r)}{\Gamma(s)}e^\tau \tau^{s-r}+\frac{\Gamma(r)}{\Gamma(r-s)} (-\tau)^{-s},\nn\\
&U(s,r,\tau)\sim \tau^{-s},
\end{align}
where the cut of $\arg \tau$ is taken to be the negative real axis and $\arg (-1)=\pi$. They are used to determine, under specified initial conditions at $t\rar -\infty$, the constants $A$ and $B$. Under the initial conditions, $\varphi_1(-\infty)=0$ and $|\varphi_2(-\infty)|=1$, we obtain $A=0$ and $|B|=e^{\varepsilon\gamma/(2\beta)-\pi k(1-\cos\theta)/4\beta}$. The asymptotic behaviors at $t\rar \infty$ read: $\varphi_1(\infty)=0$ and $|\varphi_2(\infty)|=e^{-\pi k(1-\cos\theta)/2\beta}$. Squares of moduli of $\varphi_1(\infty)$ and $\varphi_2(\infty)$ then give the transition probabilities from the second state:
\begin{align}\label{from2}
&P_{2\rar 1}=0,\nn\\
&P_{2\rar 2}=e^{-\pi k(1-\cos\theta)/\beta}.
\end{align}
Under the initial conditions, $|\varphi_1(-\infty)|=1$ and $\varphi_2(-\infty)=0$, we obtain $|A|=|\cos\frac{\theta}{2}|e^{-\varepsilon\gamma/2\beta}\sqrt{\frac{1- e^{-2\pi k/\beta}}{1-e^{-\pi k(1-\cos\theta)/\beta}}}$ and $B=-A\frac{\Gamma(r)}{\Gamma(r-s)}e^{-\pi k(1-\cos\theta)/2\beta}$. The transition probabilities from the first state then read:
\begin{align}\label{from1}
&P_{1\rar 1}=e^{-\pi k(1+\cos\theta)/\beta},\nn\\
&P_{1\rar 2}=e^{-2\varepsilon\gamma/\beta}[1-e^{-\pi k(1+\cos\theta)/\beta}][1-e^{-\pi k(1-\cos\theta)/\beta}].
\end{align}
The results \eqref{from2} and \eqref{from1} agree with Eq. \eqref{prm2} obtained via the contour deformation method. 
In addition we have obtained the non-universal element $X \equiv P_{1\rar 2}$.

\end{document}